\documentclass{PoS}
\usepackage{ifthen} 
\newboolean{pdflatex}
\setboolean{pdflatex}{true} 

\newboolean{articletitles}
\setboolean{articletitles}{true} 

\newboolean{uprightparticles}
\setboolean{uprightparticles}{false} 

\usepackage{microtype}
\usepackage{lineno}  
\usepackage{xspace} 
\usepackage{caption} 

\usepackage{graphicx}  
\usepackage{color}
\usepackage{colortbl}
\graphicspath{{./figs/}} 

\usepackage{amsmath} 
\usepackage{amssymb}
\usepackage{amsfonts}
\usepackage{upgreek} 

\newcommand*\patchAmsMathEnvironmentForLineno[1]{%
\expandafter\let\csname old#1\expandafter\endcsname\csname #1\endcsname
\expandafter\let\csname oldend#1\expandafter\endcsname\csname
end#1\endcsname
 \renewenvironment{#1}%
   {\linenomath\csname old#1\endcsname}%
   {\csname oldend#1\endcsname\endlinenomath}%
}
\newcommand*\patchBothAmsMathEnvironmentsForLineno[1]{%
  \patchAmsMathEnvironmentForLineno{#1}%
  \patchAmsMathEnvironmentForLineno{#1*}%
}
\AtBeginDocument{%
\patchBothAmsMathEnvironmentsForLineno{equation}%
\patchBothAmsMathEnvironmentsForLineno{align}%
\patchBothAmsMathEnvironmentsForLineno{flalign}%
\patchBothAmsMathEnvironmentsForLineno{alignat}%
\patchBothAmsMathEnvironmentsForLineno{gather}%
\patchBothAmsMathEnvironmentsForLineno{multline}%
\patchBothAmsMathEnvironmentsForLineno{eqnarray}%
}

\usepackage{hyperxmp}

\usepackage[colorinlistoftodos,textsize=scriptsize]{todonotes}

\usepackage[bottom,flushmargin,hang,multiple]{footmisc}

\usepackage{xspace} 
\usepackage{upgreek}

\def\babar  {\mbox{BaBar}\xspace}

\def\MagUp {\mbox{\em Mag\kern -0.05em Up}\xspace}

\ifthenelse{\boolean{uprightparticles}}%
{
 
 \def\Pgamma      {\ensuremath{\upgamma}\xspace}

 \def\Pmu         {\ensuremath{\upmu}\xspace}                 
 \def\Pnu         {\ensuremath{\upnu}\xspace}                 
                  
 \def\Ppi         {\ensuremath{\uppi}\xspace}

 \def\Ptau        {\ensuremath{\uptau}\xspace}                 
                  
 \def\Pphi        {\ensuremath{\upphi}\xspace}

 \def\Ppsi        {\ensuremath{\uppsi}\xspace}

 \def\PDelta      {\ensuremath{\Delta}\xspace}                 
 \def\PXi         {\ensuremath{\Xi}\xspace}                 
 \def\PLambda     {\ensuremath{\Lambda}\xspace}                 
 \def\PSigma      {\ensuremath{\Sigma}\xspace}                 
 \def\POmega      {\ensuremath{\Omega}\xspace}                 
 \def\PUpsilon    {\ensuremath{\Upsilon}\xspace}

 \def\PB      {\ensuremath{\mathrm{B}}\xspace}                 
 \def\PC      {\ensuremath{\mathrm{C}}\xspace}                 
 \def\PD      {\ensuremath{\mathrm{D}}\xspace}

 \def\PJ      {\ensuremath{\mathrm{J}}\xspace}                 
 \def\PK      {\ensuremath{\mathrm{K}}\xspace}

 \def\Pb      {\ensuremath{\mathrm{b}}\xspace}                 
 \def\Pc      {\ensuremath{\mathrm{c}}\xspace}                 
 \def\Pd      {\ensuremath{\mathrm{d}}\xspace}                 
 \def\Pe      {\ensuremath{\mathrm{e}}\xspace}

 \def\Pi      {\ensuremath{\mathrm{i}}\xspace}

 \def\Pp      {\ensuremath{\mathrm{p}}\xspace}                 
 \def\Pq      {\ensuremath{\mathrm{q}}\xspace}                 
                  
 \def\Ps      {\ensuremath{\mathrm{s}}\xspace}                 
 \def\Pt      {\ensuremath{\mathrm{t}}\xspace}                 
 \def\Pu      {\ensuremath{\mathrm{u}}\xspace}

 \def\thebaroffset{0.0em}
}
{
 
 \def\Pgamma      {\ensuremath{\gamma}\xspace}

 \def\Pmu         {\ensuremath{\mu}\xspace}                 
 \def\Pnu         {\ensuremath{\nu}\xspace}                 
                  
 \def\Ppi         {\ensuremath{\pi}\xspace}

 \def\Ptau        {\ensuremath{\tau}\xspace}                 
                  
 \def\Pphi        {\ensuremath{\phi}\xspace}

 \def\Ppsi        {\ensuremath{\psi}\xspace}                 
                  
 \mathchardef\PDelta="7101
 \mathchardef\PXi="7104
 \mathchardef\PLambda="7103
 \mathchardef\PSigma="7106
 \mathchardef\POmega="710A
 \mathchardef\PUpsilon="7107
                  
 \def\PB      {\ensuremath{B}\xspace}                 
 \def\PC      {\ensuremath{C}\xspace}                 
 \def\PD      {\ensuremath{D}\xspace}

 \def\PJ      {\ensuremath{J}\xspace}                 
 \def\PK      {\ensuremath{K}\xspace}

 \def\Pb      {\ensuremath{b}\xspace}                 
 \def\Pc      {\ensuremath{c}\xspace}                 
 \def\Pd      {\ensuremath{d}\xspace}                 
 \def\Pe      {\ensuremath{e}\xspace}

 \def\Pi      {\ensuremath{i}\xspace}

 \def\Pp      {\ensuremath{p}\xspace}                 
 \def\Pq      {\ensuremath{q}\xspace}                 
                  
 \def\Ps      {\ensuremath{s}\xspace}                 
 \def\Pt      {\ensuremath{t}\xspace}                 
 \def\Pu      {\ensuremath{u}\xspace}

 \def\thebaroffset{0.18em}
}
\newcommand{\offsetoverline}[2][\thebaroffset]{\kern #1\overline{\kern -#1 #2}}%

\makeatletter
\ifcase \@ptsize \relax
  \newcommand{\miniscule}{\@setfontsize\miniscule{4}{5}}
\or
  \newcommand{\miniscule}{\@setfontsize\miniscule{5}{6}}
\or
  \newcommand{\miniscule}{\@setfontsize\miniscule{5}{6}}
\fi
\makeatother

\DeclareRobustCommand{\optbar}[1]{\shortstack{{\miniscule (\rule[.5ex]{1.25em}{.18mm})}
  \\ [-.7ex] $#1$}}

\def\epm        {{\ensuremath{\Pe^\pm}}\xspace} 
 
\def\epem       {{\ensuremath{\Pe^+\Pe^-}}\xspace}

\def\muon       {{\ensuremath{\Pmu}}\xspace}
\def\mup        {{\ensuremath{\Pmu^+}}\xspace}
\def\mun        {{\ensuremath{\Pmu^-}}\xspace} 
 
\def\mump       {{\ensuremath{\Pmu^\mp}}\xspace} 
\def\mumu       {{\ensuremath{\Pmu^+\Pmu^-}}\xspace}

\def\taum       {{\ensuremath{\Ptau^-}}\xspace}

\def\ellp       {{\ensuremath{\ell^+}}\xspace}
\def\ellell     {\ensuremath{\ell^+ \ell^-}\xspace}

\def\neu        {{\ensuremath{\Pnu}}\xspace}
\def\neub       {{\ensuremath{\overline{\Pnu}}}\xspace}

\def\g      {{\ensuremath{\Pgamma}}\xspace}

\def\quark     {{\ensuremath{\Pq}}\xspace}

\def\uquark    {{\ensuremath{\Pu}}\xspace}

\def\dquark    {{\ensuremath{\Pd}}\xspace}

\def\squark    {{\ensuremath{\Ps}}\xspace}

\def\cquark    {{\ensuremath{\Pc}}\xspace}

\def\bquark    {{\ensuremath{\Pb}}\xspace}

\def\tquark    {{\ensuremath{\Pt}}\xspace}

\def\pion   {{\ensuremath{\Ppi}}\xspace}

\def\pip    {{\ensuremath{\pion^+}}\xspace}
\def\pim    {{\ensuremath{\pion^-}}\xspace}

\def\pimp   {{\ensuremath{\pion^\mp}}\xspace}

\def\kaon    {{\ensuremath{\PK}}\xspace}

\def\KorKbar {\kern \thebaroffset\optbar{\kern -\thebaroffset \PK}{}\xspace}
\def\Kz      {{\ensuremath{\kaon^0}}\xspace}

\def\Kp      {{\ensuremath{\kaon^+}}\xspace}
\def\Km      {{\ensuremath{\kaon^-}}\xspace}
\def\Kpm     {{\ensuremath{\kaon^\pm}}\xspace}

\def\KS      {{\ensuremath{\kaon^0_{\mathrm{S}}}}\xspace}

\def\Kstarz  {{\ensuremath{\kaon^{*0}}}\xspace}

\def\Kstar   {{\ensuremath{\kaon^*}}\xspace}

\def\Kstarp  {{\ensuremath{\kaon^{*+}}}\xspace}

\def\Dbar    {{\ensuremath{\offsetoverline{\PD}}}\xspace}
\def\D       {{\ensuremath{\PD}}\xspace}

\def\DorDbar {\kern \thebaroffset\optbar{\kern -\thebaroffset \PD}\xspace}
\def\Dz      {{\ensuremath{\D^0}}\xspace}
\def\Dzb     {{\ensuremath{\Dbar{}^0}}\xspace}
\def\Dp      {{\ensuremath{\D^+}}\xspace}
\def\Dm      {{\ensuremath{\D^-}}\xspace}

\def\DpDm    {\ensuremath{\Dp {\kern -0.16em \Dm}}\xspace}
\def\Dstar   {{\ensuremath{\D^*}}\xspace}

\def\B       {{\ensuremath{\PB}}\xspace}
\def\Bbar    {{\ensuremath{\offsetoverline{\PB}}}\xspace}

\def\BorBbar {\kern \thebaroffset\optbar{\kern -\thebaroffset \PB}\xspace}
\def\Bz      {{\ensuremath{\B^0}}\xspace}

\def\Bd      {{\ensuremath{\B^0}}\xspace}

\def\BdorBdbar {\kern \thebaroffset\optbar{\kern -\thebaroffset \Bd}\xspace}
\def\Bu      {{\ensuremath{\B^+}}\xspace}
\def\Bub     {{\ensuremath{\B^-}}\xspace}
\def\Bp      {{\ensuremath{\Bu}}\xspace}
\def\Bm      {{\ensuremath{\Bub}}\xspace}

\def\Bs      {{\ensuremath{\B^0_\squark}}\xspace}

\def\BsorBsbar {\kern \thebaroffset\optbar{\kern -\thebaroffset \Bs}\xspace}
\def\Bc      {{\ensuremath{\B_\cquark^+}}\xspace}

\def\jpsi     {{\ensuremath{{\PJ\mskip -3mu/\mskip -2mu\Ppsi}}}\xspace}

\def\Y#1S{\ensuremath{\PUpsilon{(#1S)}}\xspace}

\def\proton      {{\ensuremath{\Pp}}\xspace}

\def\Lz          {{\ensuremath{\PLambda}}\xspace}
\def\Lbar        {{\ensuremath{\offsetoverline{\PLambda}}}\xspace}
\def\LorLbar     {\kern \thebaroffset\optbar{\kern -\thebaroffset \PLambda}\xspace}

\def\Sigmares    {{\ensuremath{\PSigma}}\xspace}

\def\Xires       {{\ensuremath{\PXi}}\xspace}

\def\Omegares    {{\ensuremath{\POmega}}\xspace}

\def\Lc          {{\ensuremath{\Lz^+_\cquark}}\xspace}

\def\Xicp        {{\ensuremath{\Xires^+_\cquark}}\xspace}

\def\Omegac      {{\ensuremath{\Omegares^0_\cquark}}\xspace}

\def\Xiccp       {{\ensuremath{\Xires^+_{\cquark\cquark}}}\xspace}
\def\Xiccpp      {{\ensuremath{\Xires^{++}_{\cquark\cquark}}}\xspace}

\def\Lb           {{\ensuremath{\Lz^0_\bquark}}\xspace}

\def\Sigmabz      {{\ensuremath{\Sigmares_\bquark^0}}\xspace}

\def\Omegab       {{\ensuremath{\Omegares^-_\bquark}}\xspace}

\def\BF         {{\ensuremath{\mathcal{B}}}\xspace}

\newcommand{\decay}[2]{\ensuremath{#1\!\to #2}\xspace} 

\def\to                 {\ensuremath{\rightarrow}\xspace}

\def\qsq       {{\ensuremath{q^2}}\xspace}

\def\CP                {{\ensuremath{C\!P}}\xspace}

\def\Vub  {{\ensuremath{V_{\uquark\bquark}}}\xspace}
\def\Vcb  {{\ensuremath{V_{\cquark\bquark}}}\xspace}
\def\Vtb  {{\ensuremath{V_{\tquark\bquark}}}\xspace}

\def\Vtss  {{\ensuremath{V_{\tquark\squark}^\ast}}\xspace}

\newcommand{\phis}{{\ensuremath{\phi_{\squark}}}\xspace}

\def\AT#1     {\ensuremath{A_{\mathrm{T}}^{#1}}\xspace}           

\def\Bsmm     {\decay{\Bs}{\mup\mun}}
\def\Bdmm     {\decay{\Bd}{\mup\mun}}

\def\C#1      {\ensuremath{\mathcal{C}_{#1}}\xspace}                       
\def\Cp#1     {\ensuremath{\mathcal{C}_{#1}^{'}}\xspace}                    
\def\Ceff#1   {\ensuremath{\mathcal{C}_{#1}^{\mathrm{(eff)}}}\xspace}        
\def\Cpeff#1  {\ensuremath{\mathcal{C}_{#1}^{'\mathrm{(eff)}}}\xspace}       
\def\Ope#1    {\ensuremath{\mathcal{O}_{#1}}\xspace}                       
\def\Opep#1   {\ensuremath{\mathcal{O}_{#1}^{'}}\xspace}                    

\newcommand{\aunit}[1]{\ensuremath{\text{\,#1}}}       

\newcommand{\tev}{\aunit{Te\kern -0.1em V}\xspace}
\newcommand{\gev}{\aunit{Ge\kern -0.1em V}\xspace}
\newcommand{\mev}{\aunit{Me\kern -0.1em V}\xspace}
\newcommand{\kev}{\aunit{ke\kern -0.1em V}\xspace}
\newcommand{\ev}{\aunit{e\kern -0.1em V}\xspace}
 
\newcommand{\mevc}{\ensuremath{\aunit{Me\kern -0.1em V\!/}c}\xspace}
\newcommand{\gevc}{\ensuremath{\aunit{Ge\kern -0.1em V\!/}c}\xspace}
\newcommand{\mevcc}{\ensuremath{\aunit{Me\kern -0.1em V\!/}c^2}\xspace}
\newcommand{\gevcc}{\ensuremath{\aunit{Ge\kern -0.1em V\!/}c^2}\xspace}
\newcommand{\gevgevcccc}{\ensuremath{\gev^2\!/c^4}\xspace} 

\def\km   {\aunit{km}\xspace}

\def\pb {\aunit{pb}\xspace}
\def\invpb {\ensuremath{\pb^{-1}}\xspace}
\def\fb   {\ensuremath{\aunit{fb}}\xspace}
\def\invfb   {\ensuremath{\fb^{-1}}\xspace}
\def\ab   {\ensuremath{\aunit{ab}}\xspace}
\def\invab   {\ensuremath{\ab^{-1}}\xspace}

\def\fs   {\aunit{fs}}

\def\gsim{{~\raise.15em\hbox{$>$}\kern-.85em
          \lower.35em\hbox{$\sim$}~}\xspace}
\def\lsim{{~\raise.15em\hbox{$<$}\kern-.85em
          \lower.35em\hbox{$\sim$}~}\xspace}

\def\tell1  {TELL1\xspace}
\def\ukl1   {UKL1\xspace}

\usepackage{cite} 
\usepackage{mciteplus}

\usepackage{decays}
\usepackage{url}
\usepackage{hyperxmp}

\newcolumntype{C}{>{$}c<{$}}
\newcolumntype{R}{>{$}r<{$}}
\newcolumntype{L}{>{$}l<{$}}
\newcommand{\IF}[4]{\ifthenelse{\equal{#1}{#2}}{#3}{#4}}%
\renewcommand{\PC}[1][none]{{\ensuremath{\IF{#1}{none}{P_c^+}{P_c(#1)^+}}}\xspace}
\newcommand{\aerr}[2]{{\:}^{+{\:}#1}_{-{\:}#2}}%

\title{Beauty 2019 --- Conference summary}

\ShortTitle{Conference summary}

\author{\speaker{Patrick Koppenburg}\\
  Nikhef, Amsterdam, Netherlands\\
  E-mail: \email{patrick.koppenburg@cern.ch}}

\abstract{Some highlights from the 18$^{\rm th}$ international conference on \B physics at frontier machines are presented, including first results from the full LHC Run 2 data and from early Belle II data. }

\FullConference{18th International Conference on \B-Physics at Frontier Machines --- Beauty 2019 ---\\
		29 September / 4 October, 2019\\
		Ljubljana, Slovenia}

\begin{document}

\section{Introduction}
Beauty 2019 was a lively conference, where we saw first results from the full
LHC Run~2 data and from early Belle 2 data.
There was excitement due to new announcements, mostly in spectroscopy, and
in anticipation of updates of the now long-standing flavour anomalies.
The observation of \CP violation in charm --- certainly the HEP highlight of 2019 ---
also triggered many discussions.
But the main initial shock was the announcement of the bankruptcy of the
local carrier Adria airways,
which forced most of the attendees to rearrange their travel plans at the last minute.

The scene was nicely set by Chris Quigg, who asked 50 questions to be answered by
experiment and theory~\cite{B:Quigg}.

\section{Spectroscopy}\label{Sec:Spectroscopy}
Who would have guessed that doubly charmed baryons would feature prominently at a Beauty physics conference?
Fifteen years after the putative observation of the \Xiccp baryon 
at a mass of $3519\mevcc$ by the SELEX~\cite{Mattson:2002vu,*Ocherashvili:2004hi} experiment,
LHCb observed its doubly charged counterpart \Xiccpp~\cite{LHCb-PAPER-2017-018,B:Needham}.
It was seen in the decay modes \decay{\Xiccpp}{\Lc\Km\pip\pip} and
\decay{\Xiccpp}{\Xires_c^+\pip}~\cite{LHCb-PAPER-2018-026}, but not in 
\decay{\Xiccpp}{\Dp p\Km\pip}~\cite{LHCb-PAPER-2019-011}.
In the meantime we know its mass ($3621.24\pm0.65\pm0.31\mev$)~\cite{LHCb-PAPER-2019-037};
its lifetime ($256\aerr{24}{22}\pm14\fs$)~\cite{LHCb-PAPER-2018-019},
which confirms it as a weakly decaying particle; 
and its production rate in $pp$ collisions
at $\sqrt{s}=13\tev$~\cite{LHCb-PAPER-2019-035}.

\begin{figure}[b]
  \includegraphics[height=0.27\textwidth]{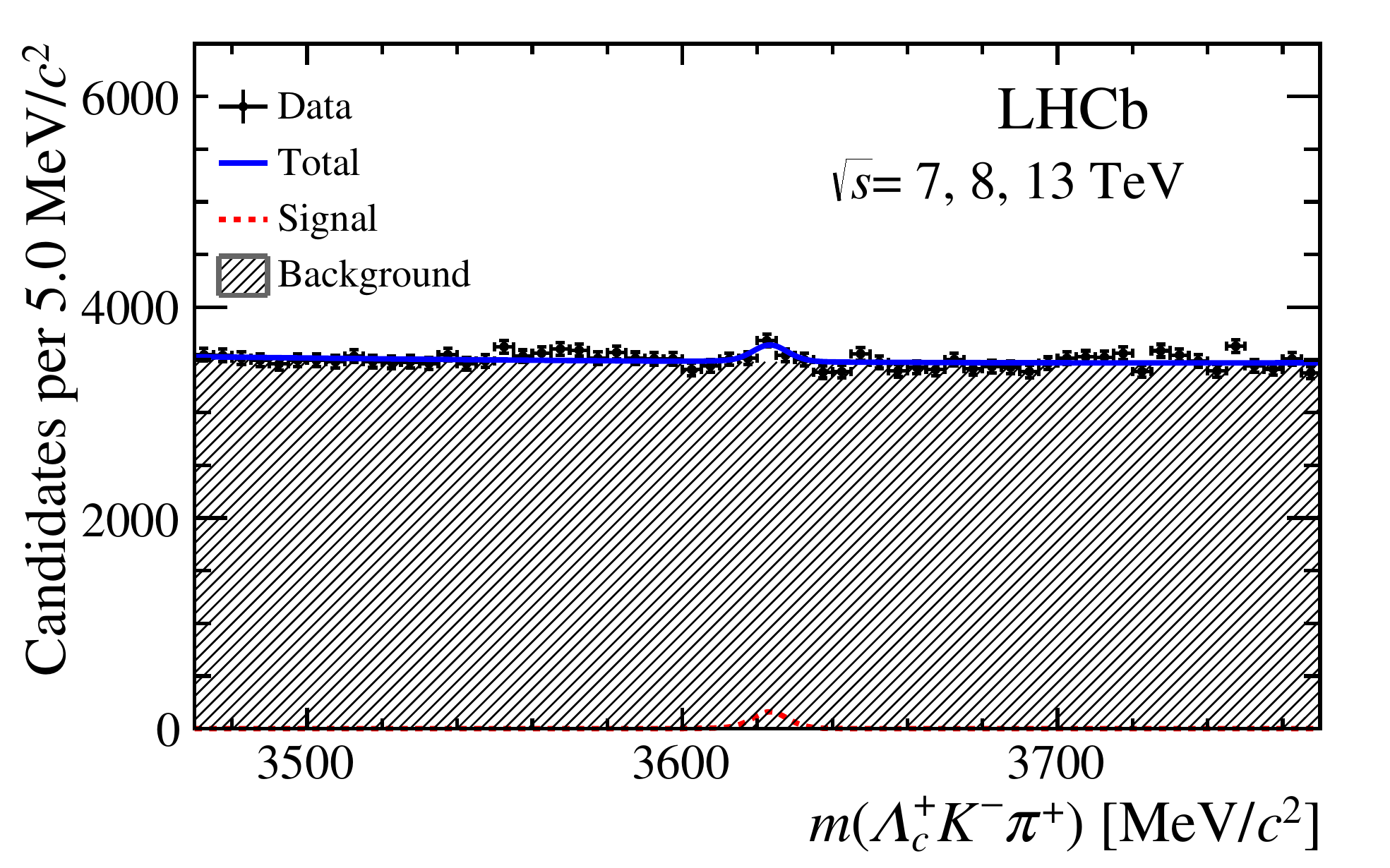}\quad
  \includegraphics[height=0.27\textwidth]{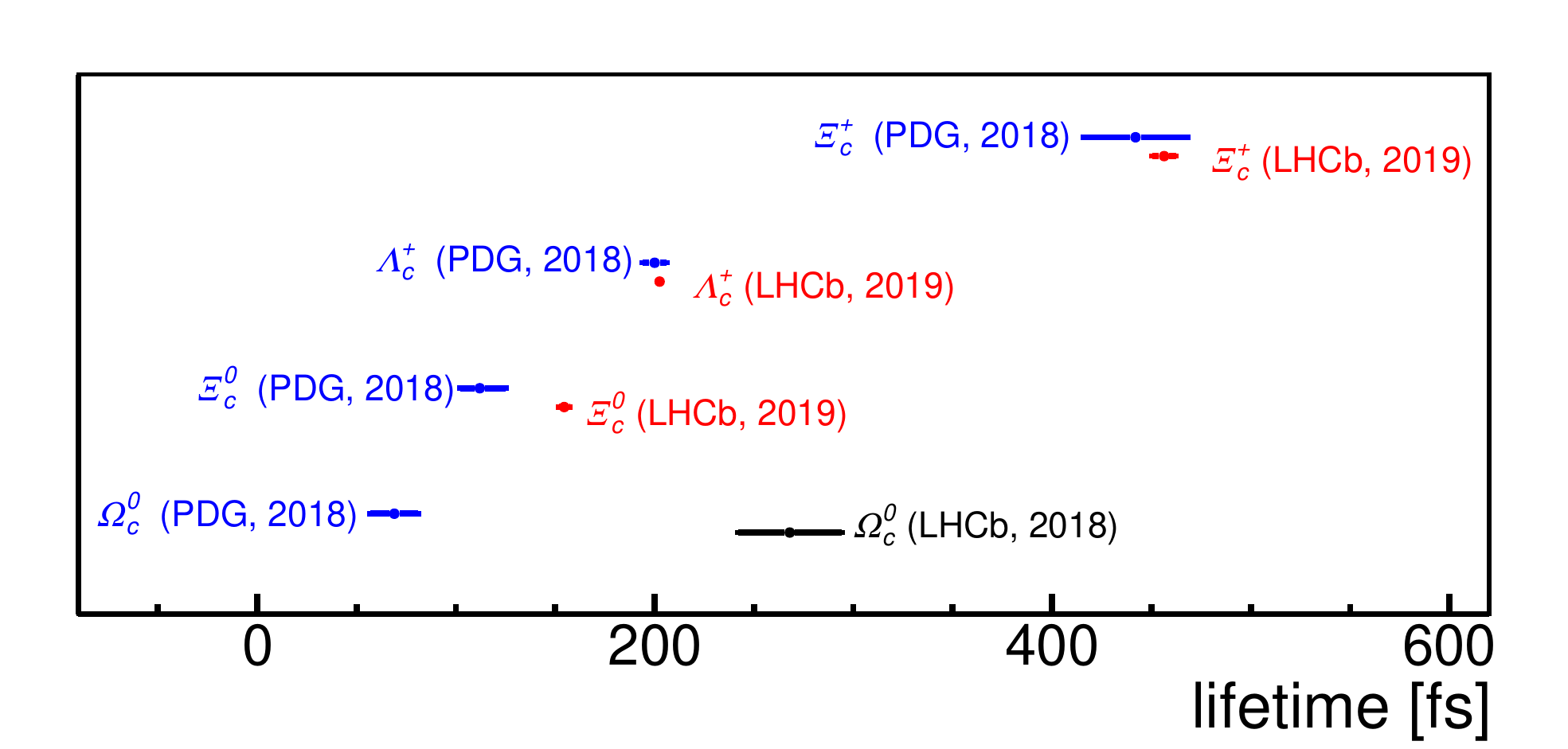}
  \caption{(left) Mass spectrum of $\Lc$ in the search for the \Xiccp baryon at LHCb~\cite{LHCb-PAPER-2019-029}. (right) Lifetimes of charm baryons~\cite{LHCb-PAPER-2019-008}.}\label{Fig:cbaryons}
\end{figure}

However, the mass measurement poses a problem: the \Xiccp and \Xiccpp states
cannot be isospin partners as their masses differ by 100\mevcc, while at most 1\mev is expected.
LHCb thus started looking for the singly charged state and did not find any excess at $3519\mevcc$~\cite{LHCb-PAPER-2019-029}.
However, there is a $2.7\sigma$ excess at a mass of $3621\mevcc$ (Fig.~\ref{Fig:cbaryons}), which
is the near the expected mass for an isospin partner of the 
\Xiccpp baryon~\cite{B:Prelovsek,*Padmanath:2019wid}. More data  will tell if the
SELEX state is a statistical fluctuation,
if there is an unusual ispospin splitting, or if the two states
are different in nature.

Another conundrum is the \Omegac lifetime:
the FOCUS, WA89 and E687 experiments give a combined lifetime of
$69\pm12\fs$~\cite{Link:2003nq,*Adamovich:1995pf,*Frabetti:1995bi,PDG2018}.
It is to be noted here that these fixed-target experiments report lifetimes at the edge
of their resolution of 50 to 70\fs. With a much larger sample,
LHCb get $284\pm25\fs$~\cite{LHCb-PAPER-2018-028},
which now places the \Omegac lifetime in between those of the \Lc and \Xicp
baryons~\cite{LHCb-PAPER-2019-008}, as shown in Fig.~\ref{Fig:cbaryons}.

The recent observation of excited \Bc states nicely demonstrates the importance of resolution and large data samples.
ATLAS first observed one state with 8\tev data~\cite{Aad:2014laa}, which was then recently resolved into two by CMS~\cite{Sirunyan:2019osb,*B:Cristella} and LHCb~\cite{LHCb-PAPER-2019-007,B:Needham}.
Owing to the larger sample only CMS can claim the observation of both states, while LHCb has the better resolution on their masses.

More new particles have been announced by LHCb.
A new $X(3842)$ meson seen in the \Dp\Dm and \Dz\Dzb spectra could be the spin-3 $\psi_3(1^3D_3)$ state~\cite{B:Needham,LHCb-PAPER-2019-005}.
Also, two resonances appear in the \Lb{}\pip{}\pim spectrum~\cite{LHCb-PAPER-2019-025,B:Needham}.
The tentatively called $\Lz_b(6152)^0$ baryon decays to $\Sigmares_b^\pm\pimp$ and $\Sigmares_b^{*\pm}\pimp$ while the lighter $\Lz_b(6146)^0$ decays only to $\Sigmares_b^{*\pm}\pimp$.
They are likely to be the $\Lz_b(1D)^0$ doublet with $J^P=\frac{3}{2}^+$ and $J^P=\frac{5}{2}^+$.
They could also be excited \Sigmabz baryons, but this hypothesis is disfavoured~\cite{Chen:2019ywy,*Liang:2019aag}.
A much broader \Lb{}\pip{}\pim resonance, consistent with being the $\Lz_b(2S)^0$ state, was later reported using the same dataset~\cite{LHCb-PAPER-2019-045}. These states are also seen at the CMS experiment~\cite{Sirunyan:2020gtz}.

\begin{figure}[t]\centering
  \includegraphics[height=0.40\textwidth]{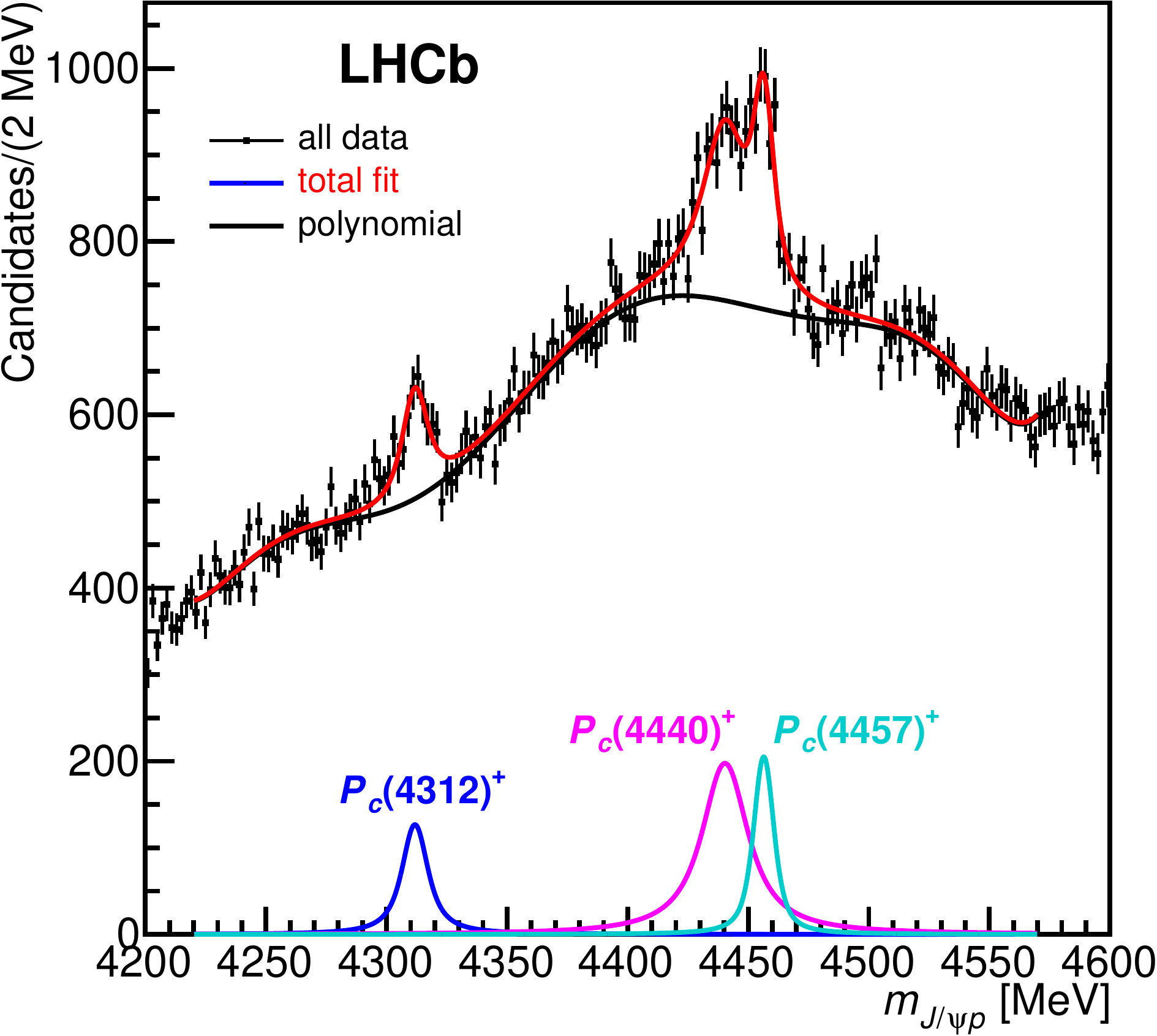}\quad
  \includegraphics[height=0.40\textwidth]{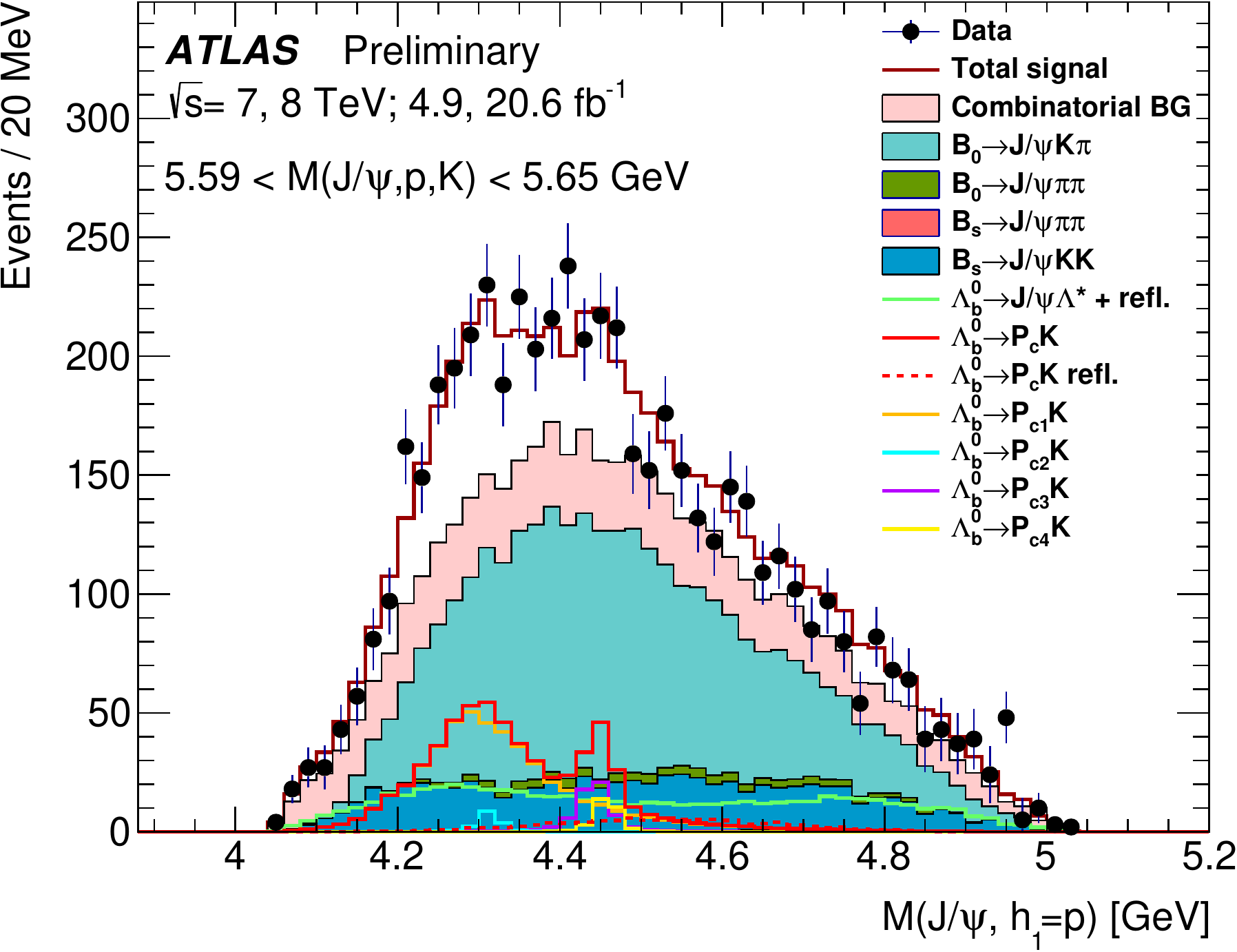}
  \caption{Fits to the \jpsi\proton mass spectrum in \decay{\Lb}{\jpsi\proton\Km} decays at (left) LHCb~\cite{LHCb-PAPER-2019-014} and (right) ATLAS~\cite{ATLAS-CONF-2019-048}.
}\label{Fig:Pc}
\end{figure}

Finally an analysis of Run 2 \decay{\Lb}{\jpsi\proton\Km} data has unveiled additional pentaquarks: A new \PC[4312] state and overlapping \PC[4440] and \PC[4457] states were reported by LHCb~\cite{LHCb-PAPER-2019-014,B:Zhang}.
The \PC[4450] state~\cite{LHCb-PAPER-2015-029}, previously reported in a full 6-dimensional amplitude analysis, is thus split into two overlapping states, while the one-dimensional fit of Ref.~\cite{LHCb-PAPER-2019-014} has no sensitivity to the broad \PC[4380] state (Fig.~\ref{Fig:Pc}).
A 6-dimensional analysis will provide a clearer picture.
For the first time the \PC states were also confirmed by another experiment: ATLAS reported a fit including the LHCb states~\cite{ATLAS-CONF-2019-048,*B:Yeletskikh} (Fig.~\ref{Fig:Pc}), although they are not able to fully exclude the hypothesis of no pentaquarks ($p\sim9\times10^{-3}$).
Meanwhile the GlueX experiment reported no evidence for \PC states in \jpsi photoproduction~\cite{Ali:2019lzf}.
A similar study from CMS is eagerly awaited. In the meantime CMS observed the promising decay  \decay{\Bs}{\jpsi\Lz\phi} and studied the decay \decay{\Bp}{\jpsi\Lbar\proton} but do not need exotic contributions to explain the data~\cite{Sirunyan:2019wwr,*Sirunyan:2019dwp,*B:Polikarpov}.

Counting the excited \Omegab states~\cite{LHCb-PAPER-2019-042}
reported after the conference,
the LHC has now observed 33 new hadrons~\cite{NewResonances}.

\section{\boldmath\CP violation}
The most exciting \CP-violation result of the year is the observation of \CP violation in the charm sector.
Using the data collected so far, LHCb measures a significant difference $\Delta A_\CP$ between the \CP asymmetries in \decay{\Dz}{\Kp\Km} and \decay{\Dz}{\pip\pim}~\cite{LHCb-PAPER-2019-006,*B:Schubiger}.
This comes 35, 17 and 6 years after the first observation of \CP violation in the kaon, \Bd and \Bs~\cite{Christenson:1964fg,*Aubert:2001nu,*Abe:2001xe,*LHCb-PAPER-2013-018} systems, respectively.
However it is hard to tell if the measured \CP asymmetry is consistent with the SM or not~\cite{B:Nierste}.
The understanding of the SM contributions to this asymmetry has improved recently --- also thanks to wildly varying values of $\Delta A_\CP$ reported in previous measurements 
--- but the jury is still out on whether the measured value is consistent with expectations.
\begin{figure}[b]\centering
  \includegraphics[height=0.36\textwidth]{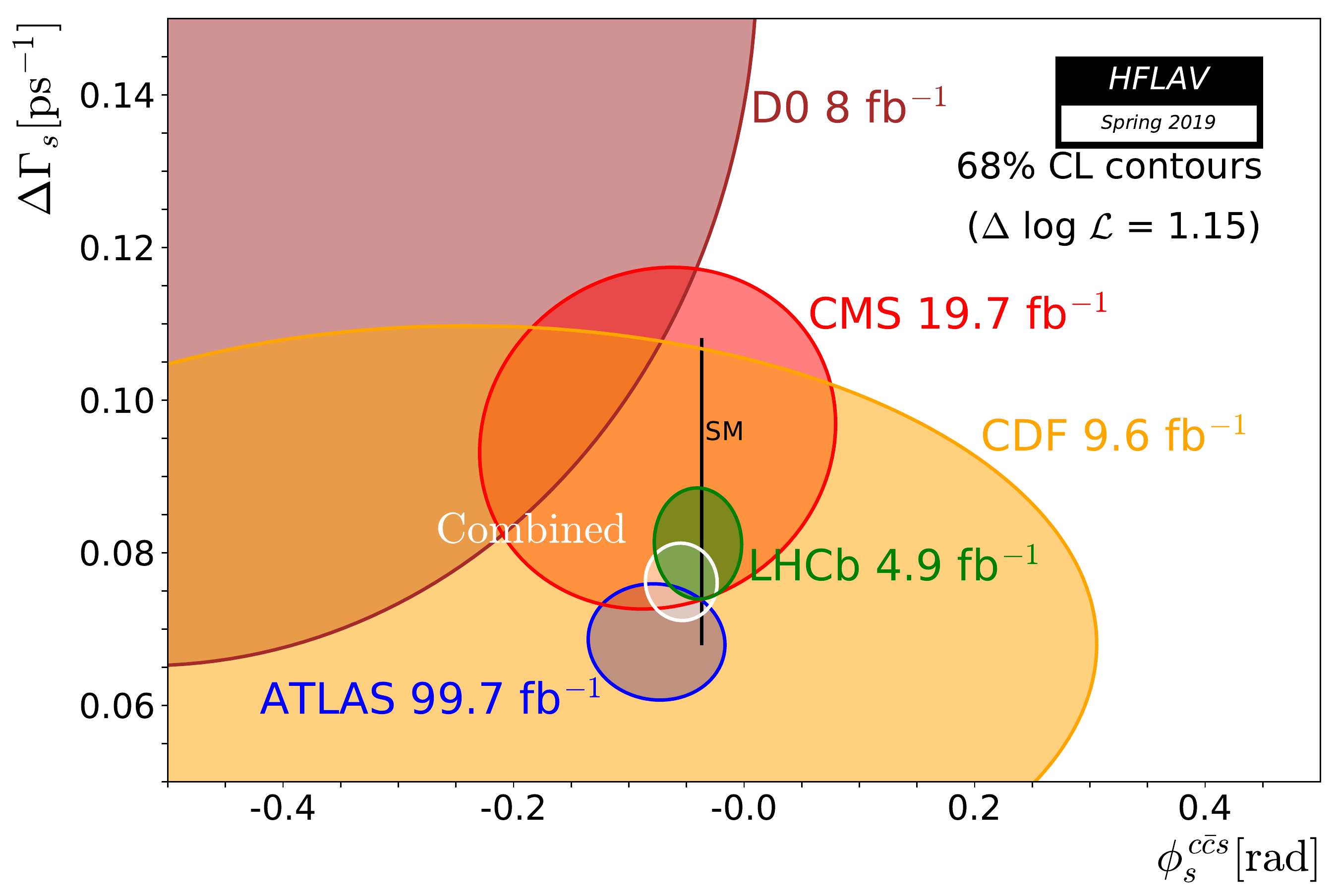}~
  \includegraphics[height=0.36\textwidth]{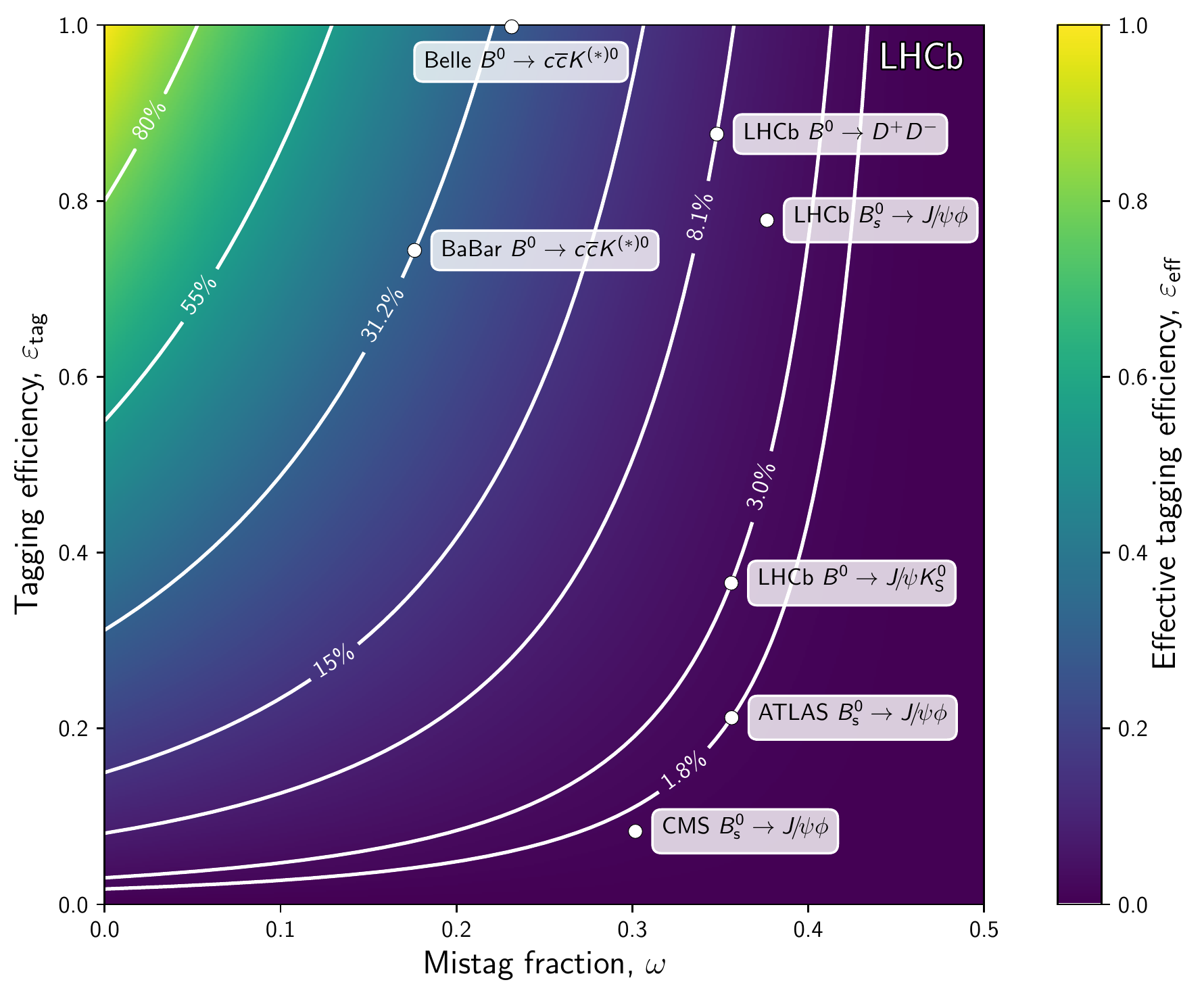}
  \caption{
    (left) Best fit of \Bs mixing parameters~\cite{HFLAV18} and
    (right) comparison of tagging efficiency and mistag fraction for selected
    \CP violation measurements~\cite{LHCb-FIGURE-2020-002,*LHCb-PAPER-2016-037,*Khachatryan:2015nza,*Aubert:2009aw,*Kakuno:2004cf,LHCb-PAPER-2015-004,LHCb-PAPER-2019-013,Aad:2020jfw}.}\label{Fig:DACP}\label{Fig:phis}\label{Fig:Tagging}
\end{figure}

In parallel there has been progress on precision measurement of \CP asymmetries
in \B decays.
The weak \Bs mixing phase, \phis, was measured with similar precision
by ATLAS and LHCb~\cite{LHCb-PAPER-2019-013,Aad:2020jfw,B:Novotny}.
The latest HFLAV average~\cite{HFLAV18} is shown in Fig.~\ref{Fig:phis} and compared to the SM prediction~\cite{Artuso:2015swg}.\footnote{This average still uses the slightly different preliminary ATLAS result~\cite{ATLAS:2019akj}.} 
The increased precision is largely due to improved flavour tagging algorithms~\cite{B:Govorkova,B:Novotny,B:Palla,*B:Calvi}. 
A comparison of the tagging efficiency and mistag rates for selected analyses in five experiments is shown in Fig.~\ref{Fig:Tagging}. The improvement is particularly striking when comparing the performance of LHCb's analyses of \decay{\Bs}{\jpsi\phi} published this year~\cite{LHCb-PAPER-2019-013} and \decay{\Bd}{\jpsi\KS}, which dates from 2015~\cite{LHCb-PAPER-2015-004}.
More developments in flavour tagging are still in the pipeline.
Improved measurements of the unitarity triangle angle $\gamma$ are also to be expected in the next years~\cite{B:Bjorn}.

\begin{figure}[t]\centering
  \includegraphics[height=0.28\textwidth]{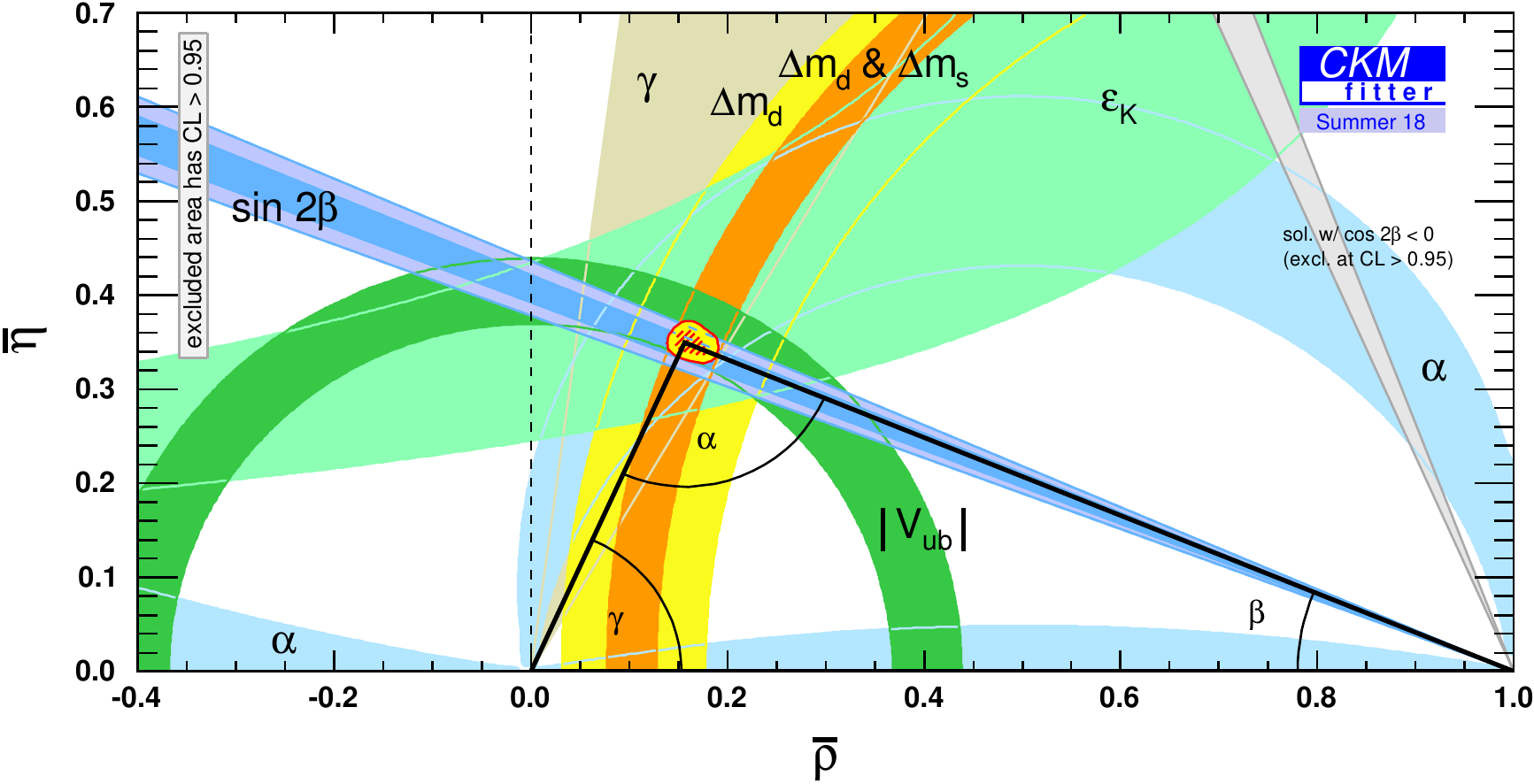}~
  \includegraphics[height=0.29\textwidth]{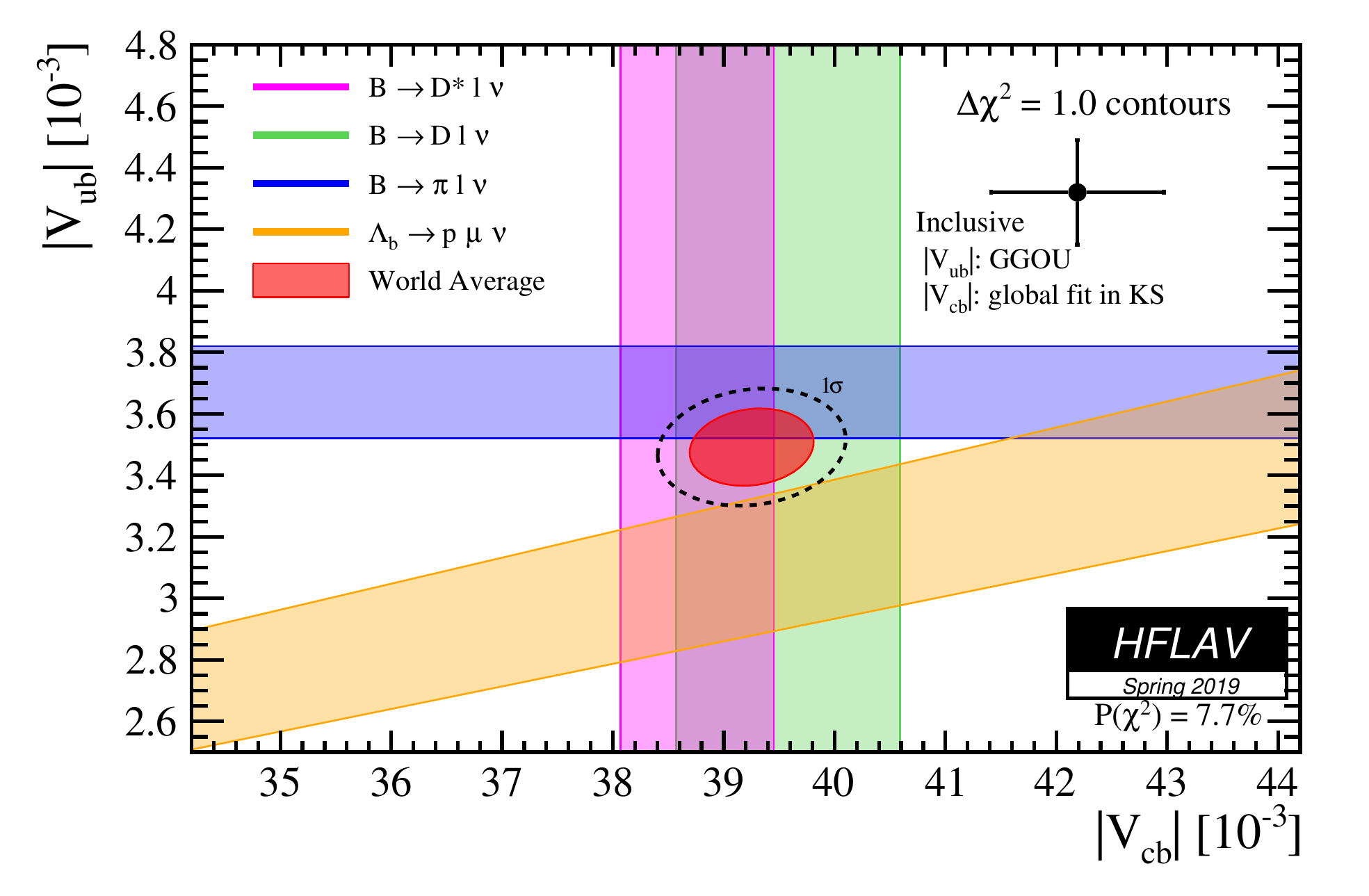}
  \caption{(left) CKM unitarity triangle fit from CKMFitter~\cite{CKMfitter2015}. (right) Determinations of $|\Vub|$ and $|\Vcb|$ using exclusive \B decays~\cite{HFLAV18}.
The average obtained from inclusive decays is shown for comparison.}\label{Fig:UT}\label{Fig:Vqb}
\end{figure}
These measurements are input to CKM unitarity triangle closure
tests~\cite{CKMfitter2015,UTfit-UT} (Fig.~\ref{Fig:UT}, left).
There is still a discrepancy in the values of the CKM matrix elements
$|\Vub|$ and $|\Vcb|$ depending on whether they are determined using inclusive
\decay{b}{q\ell\nu} or exclusive \decay{\B}{H_q\ell\nu} decays 
(Fig.~\ref{Fig:Vqb}, right).
For the latter, form factors are a critical input.
However their determination by lattice groups also varies, as shown by Witzel~\cite{B:Witzel}.
Those for the decay \decay{\Bs}{\Km\ellp\nu}~\cite{Bouchard:2014ypa,*Flynn:2015mha,*Bahr:2016ayy,*Bazavov:2019aom}, which is sensitive to $|\Vub|$, will become relevant in the near future.
The expectation from CKM fits favours the inclusive (higher) value of $|\Vcb|$ and the exclusive (lower) value of $|\Vub|$.
Recently a measurement of  \decay{\Bs}{\D_s^{(*)-}\mup\neu} by LHCb~\cite{LHCb-PAPER-2019-041} has provided a first exclusive measurement of $|\Vcb|$ using \Bs decays.
It is found to be closer to the value favoured by inclusive measurements, but also consistent with the average of exclusive determinations.
It is likely that these puzzles remain for a while.

\section{Flavour Anomalies}
\begin{figure}[b]\centering
  \includegraphics[height=0.4\textwidth]{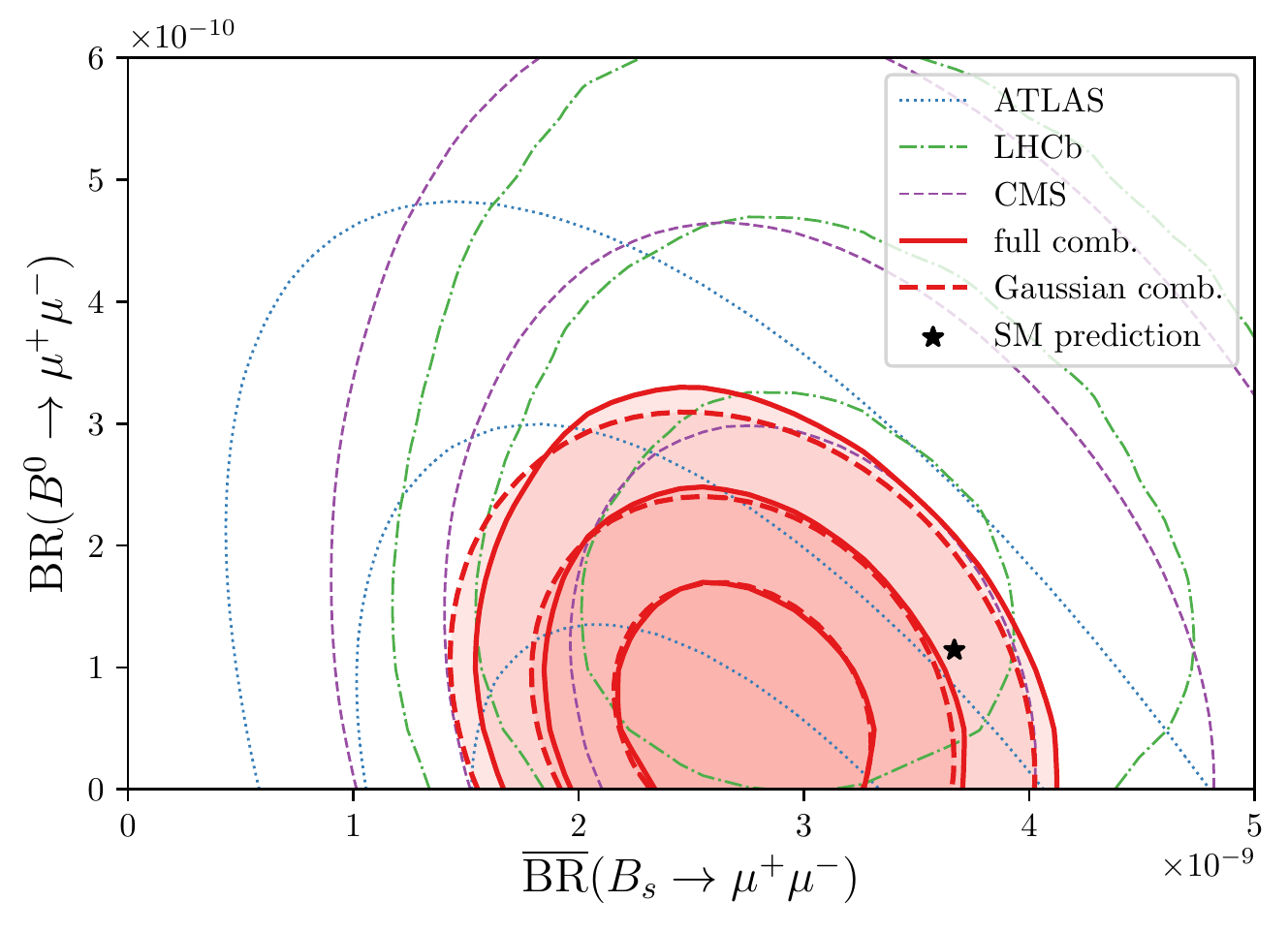}~
  \includegraphics[height=0.4\textwidth]{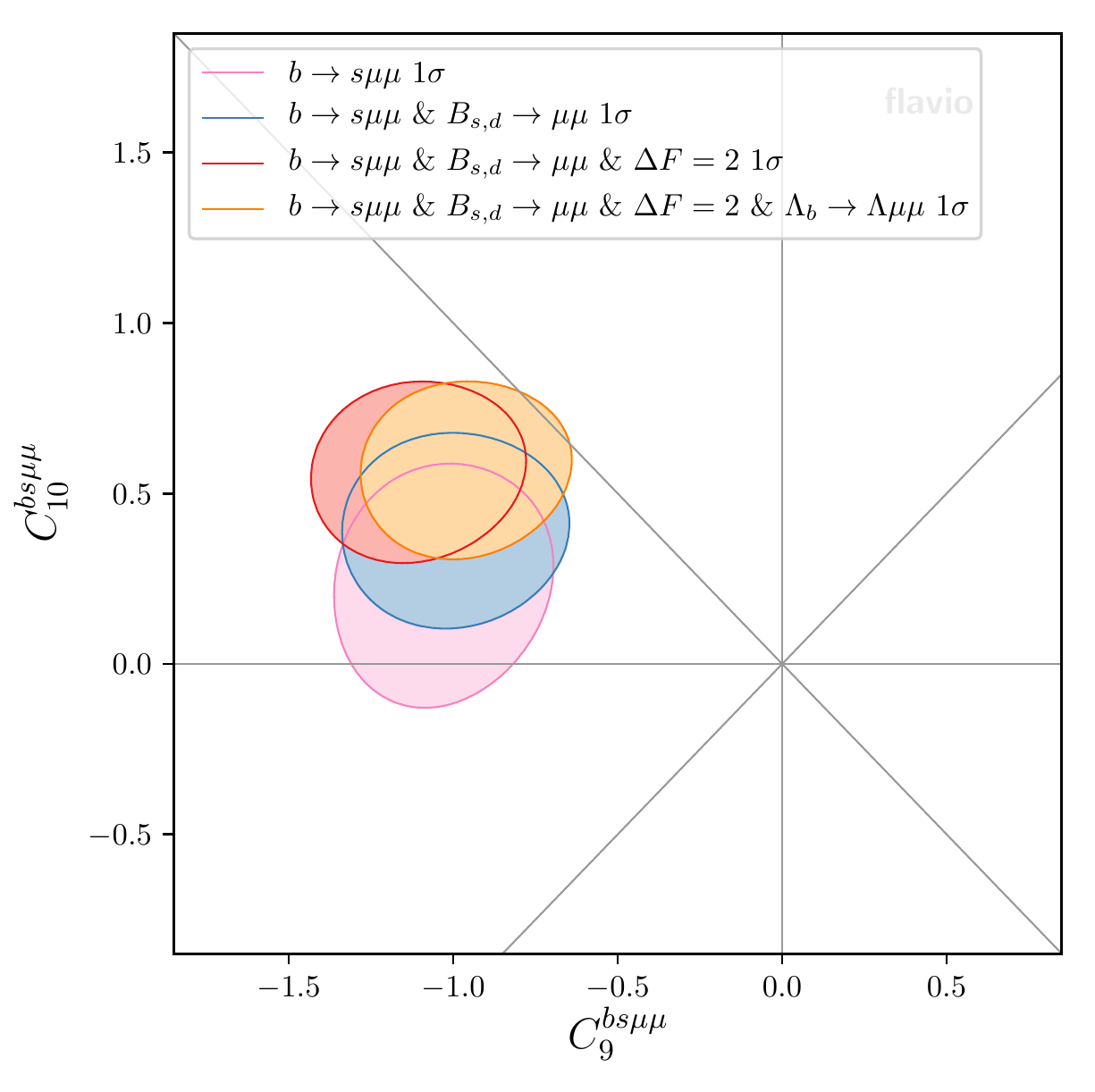}
  \caption{(left) \Bmm constraints update by Straub based on Ref.~\cite{Aebischer:2019mlg} and (right) constraints on vector and axial Wilson coefficients $C_{9\mu}^\text{NP}$ and $C_{10\mu}^\text{NP}$~\cite{Aebischer:2019mlg}.
}\label{Fig:Straub}
\end{figure}

ATLAS, CMS and LHCb have released measurements of the \Bsmm and \mbox{\Bdmm} branching fractions~\cite{LHCb-PAPER-2017-001,Aaboud:2018mst,*Sirunyan:2019xdu,*B:Walkowiak,*B:Chen} using data up to 2016.
A combination by Straub is shown in Fig.~\ref{Fig:Straub}.
It should be noted that the quoted branching fractions assume the SM value of the effective lifetime for the admixture of heavy and light \Bs states, which affects the selection efficiency~\cite{DeBruyn:2012wj}.
LHCb provides a recipe to correct for this effect~\cite{LHCb-PAPER-2017-001}.
This issue will no longer be relevant once the effective lifetime has been measured~\cite{DeBruyn:2012wk}.
LHCb and CMS have provided first measurements, although still with poor sensitivity.

It is debatable whether the \Bmm tension with the SM prediction counts as an anomaly, but 
the low rate of \bmms decays certainly does.
There is a lower rate of such decays with respect to the electronic \bees decays seen by LHCb~\cite{LHCb-PAPER-2019-040,*LHCb-PAPER-2017-013,*LHCb-PAPER-2019-009,*B:Langenbruch} and
to a lesser extent by Belle~\cite{Abdesselam:2019wac,*Abdesselam:2019lab,*B:Shun}.
Similarly, all branching fractions of \bmms decays are measured below their theoretical expectations~\cite{LHCb-PAPER-2014-006,*LHCb-PAPER-2015-009,*LHCb-PAPER-2015-023,*LHCb-PAPER-2016-012}.
Moreover, the SM prediction for \decay{\B}{\Kstar\mumu} may even be too low, as explained by Descotes-Genon~\cite{Descotes-Genon:2019bud,B:Descotes}.

The value of the $P_5'$ asymmetry in \decay{\Bz}{\Kstarz\ellell} decays~\cite{Descotes-Genon:2013vna} has been measured for muons by the LHC experiments~\cite{LHCb-PAPER-2015-051,*Aaboud:2018krd,*Sirunyan:2017dhj},\footnote{The LHCb measurement was recently updated in Ref.~\cite{LHCb-PAPER-2020-002}.} and for both muons and electrons by Belle~\cite{Wehle:2016yoi}. There are deviations from the SM predictions at the level of 2 to 3$\sigma$ in the dilepton-mass-squared region \mbox{$4<\qsq<8\gevgevcccc$}. 

\begin{table}[t]
  \centering\caption{Pulls with respect to the Standard Model of various fits to \blls data with three New Physics hypothesis~\cite{B:Descotes}.}\label{Tab:Pulls}
  \begin{tabular}{LL|CCCCCC}
  \multicolumn{2}{l}{} & \multicolumn{6}{c}{References} \\
  \multicolumn{2}{l|}{New Physics hypothesis} &
  \text{\cite{Alguero:2019ptt}} &
  \text{\cite{Aebischer:2019mlg}} &
  \text{\cite{Alok:2019ufo}} &
  \text{\cite{Arbey:2019duh}} &
  \text{\cite{DAmico:2017mtc}} &
  \text{\cite{Kowalska:2019ley}} \\ \hline
  \text{Vector:} & C_{9\mu}^\text{NP}                   & 5.6\sigma & 5.9\sigma & 6.2\sigma & 5.3\sigma & 6.5\sigma &4.7 \sigma   \\
  V-A: & C_{9\mu}^\text{NP}=-C_{10\mu}^\text{NP} & 5.2\sigma & 6.6\sigma & 6.4\sigma & 4.5\sigma & 5.9\sigma & 4.8\sigma \\
  \text{RH}: & C_{9\mu}^\text{NP}=-C_{9'\mu}^\text{NP} & 5.5\sigma &  & 6.4\sigma \\
  \end{tabular}
  \end{table}

Combining all constraints on \blls and \bgs decays, various groups perform Wilson coefficient fits which disfavour the Standard Model with large significances~\cite{Alguero:2019ptt,Aebischer:2019mlg,Alok:2019ufo,Arbey:2019duh,DAmico:2017mtc,Kowalska:2019ley}.
The pulls are listed in Table~~\ref{Tab:Pulls}, where the SM is compared to models with new vector, $V-A$, or right-handed (RH) currents. These pulls should however be taken with some care. Significant experimental signatures are eagerly awaited.

Any new physics explanation of these anomalies must leave all other
well-measured observables minimally changed.
Particularly difficult are the constraints from \Bs mixing as a new operator involving a \bquark\squark coupling would strongly affect the mixing frequency.
The measured values of $\Delta m_s$ and $\Delta m_d$ are actually a bit off the SM predictions using decay constants from the lattice~\cite{Gamiz:2009ku,*Carrasco:2013zta,*Aoki:2014nga,*Bazavov:2016nty} (by less than $2\sigma$), but the pull tends to go into the opposite direction of what would be expected from new operators required to address the flavour anomalies~\cite{DiLuzio:2019jyq,B:Vicente}.

The scale of such new physics is also very poorly constrained.
It could be as high as 30\tev for an unsuppressed coupling,
6\tev in case of CKM-suppression ($|\Vtb\Vtss|$),
2.5\tev for loop suppression ($1/16\pi^2$) and as little as
0.5\tev for both~\cite{DiLuzio:2017chi,B:Vicente}.

Any departure from lepton universality is likely associated with
some level of violation of lepton-flavour conservation.
No known symmetry principle
can protect the one in the absence of the other~\cite{Glashow:2014iga}.
It is
thus essential to look for lepton-flavour-violating decays such as \decay{\B}{Ke\mu}.
However, no hint is seen.
LHCb for instance have recently improved the limit on \decay{\Bp}{\Kp\epm\mump}~\cite{LHCb-PAPER-2019-022} to the $10^{-9}$ range.
Similarly \babar set limits in the $10^{-7}$ range for a large set of
\D decays~\cite{Lees:2019pej}.

\begin{figure}[tb]\centering
  \includegraphics[height=0.33\textwidth]{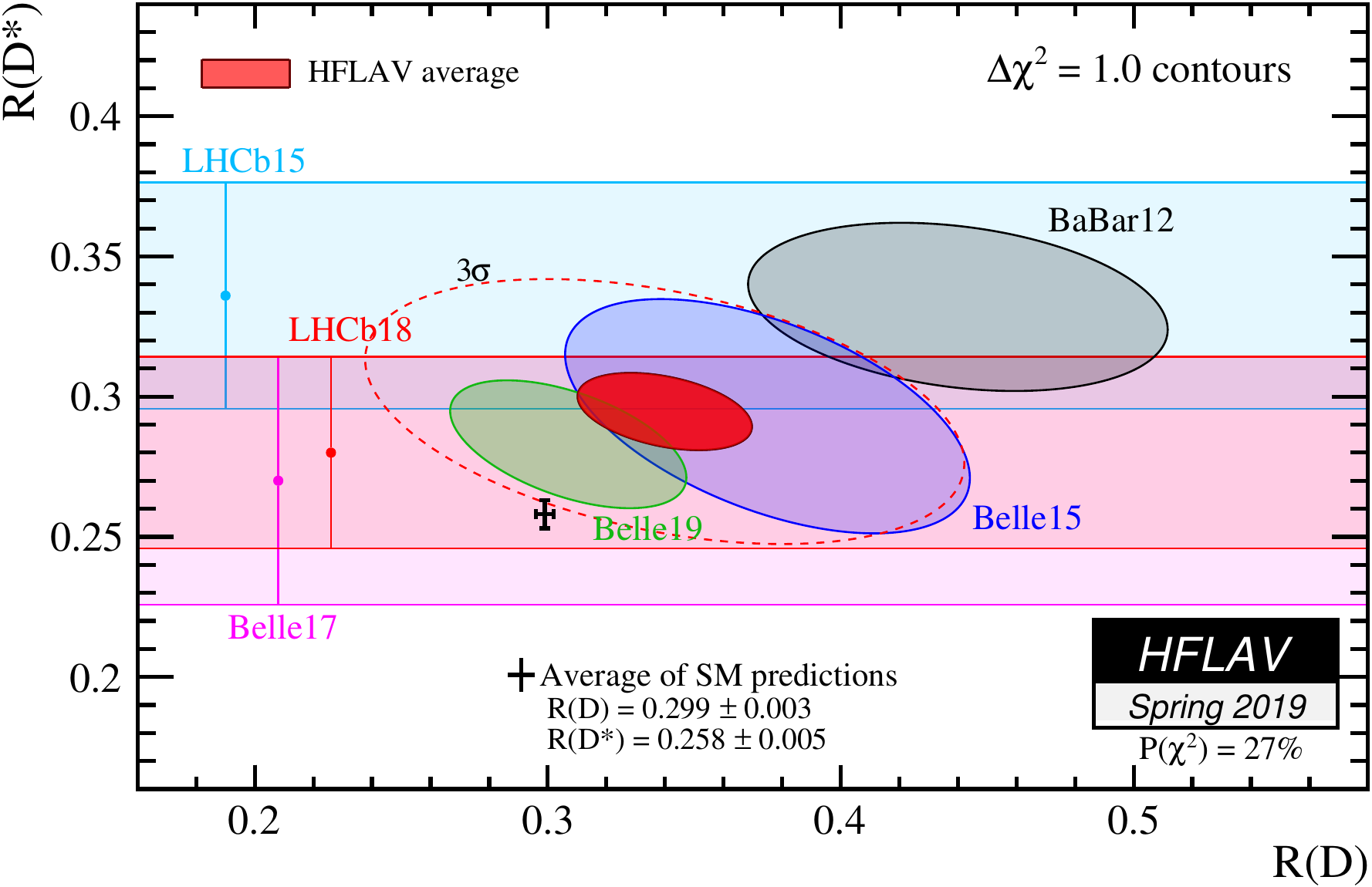}\quad
  \caption{HFLAV average of $R(D)$ and $R(\Dstar)$~\cite{HFLAV18}.
  }\label{Fig:RD}
\end{figure}

Another sign of lepton-universality violation is seen in tree-level \decay{\Bbar}{\D^{(*)}\taum\neub} decays.
Belle has presumably given their final word on the ratios
$R(D)=\BF(\decay{\Bbar}{\D\taum\neub})/\BF(\decay{\Bbar}{\D\mun\neub})$ and $R(\Dstar)$ in Ref.~\cite{Abdesselam:2019dgh}.
The HFLAV average is now $3\sigma$ away from the SM, as shown in Fig.~\ref{Fig:RD}~\cite{HFLAV18}.
\section{Outlook}
We are in exciting times for flavour physics.
Belle II, the successor of Belle, has started and is presently rediscovering rare decays such as \decay{\Bz}{\Kstarz\gamma} 
and optimising their sensitivity to \CP violation~\cite{B:Kwon,*B:Ganiev}.
They have already produced their first physics paper with 276\invpb~\cite{Adachi:2019otg}.
The aim is to collect more than 50\invab by 2027.

    Recoil mass spectrum in \decay{\epem}{\mumu Z'} search~\cite{Adachi:2019otg}
In parallel, most of the LHC Run 2 dataset is still to be analysed.
In particular the first results from the CMS parked \B sample are eagerly awaited.

In the near future, we are also expecting a much improved measurement of the muon's $g-2$ 
and improved precision in rare kaon decay experiments~\cite{B:Kucerova,*B:Pich}.

Meanwhile the theory community is working hard to meet the improved experimental precision.
Many new calculations are anticipated from the Lattice QCD collaborations, notably form factors addressing
the \Vub and \Vcb puzzles. Model builders are eagerly awaiting confirmation (or not) of the flavour anomalies
to pave the way toward a consistent New Physics model that can accommodate them.

\begin{figure}[b]\centering
  \includegraphics[width=\textwidth]{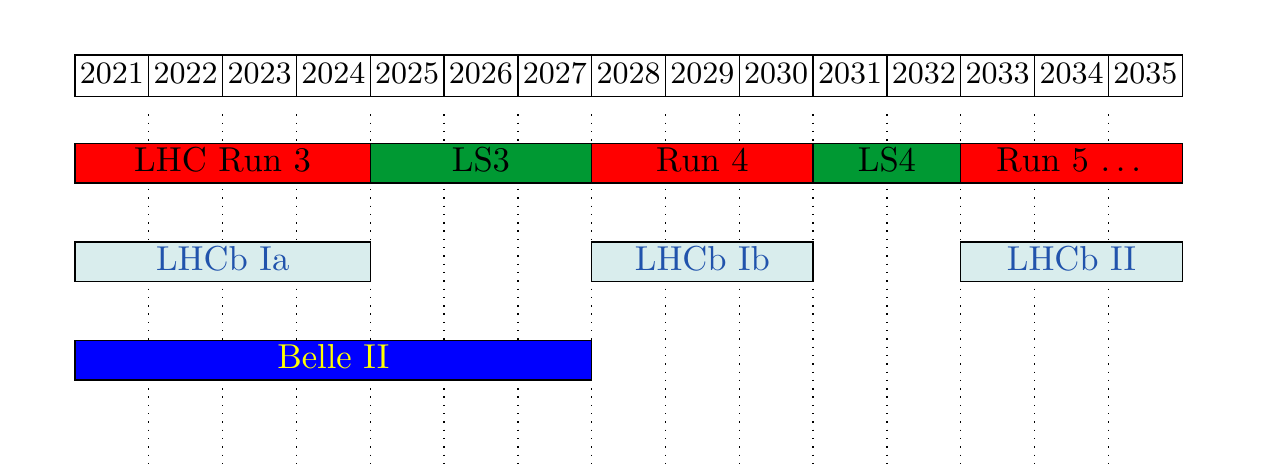}
  \caption{Timeline of LHCb and Belle II operations.}\label{Fig:TL}
\end{figure}

The LHC will resume operations in 2021.
All LHC detectors are being upgraded, but the most dramatic change is that of LHCb.
Most of the detector is being replaced in order to cope with an increased luminosity of $2\times10^{33}\rm cm^{-2}s^{-1}$ and to feed all data into a software-only trigger.
LHCb has also presented plans for a phase-II upgrade, with the plan to collect 300\invfb by the end of Run~5.
A timeline is shown in Fig.~\ref{Fig:TL}.
The expression of interest~\cite{LHCb-PII-EoI} and the physics case~\cite{LHCb-PII-Physics} were well received by the LHCC, who encourage the collaboration to present a TDR.

The future beyond that is less well defined.
The proposed 100\km FCC accelerator has also an interesting potential for flavour physics, in particular in its initial FCC-$ee$ form~\cite{Abada:2019lih,*Abada:2019zxq}.
However, it is presently hard to reply to Quigg's last question ``How do you assess the scientific potential for Beauty and in general of \dots'' followed by a list of 10 more or less realistic future projects~\cite{B:Quigg}. The answer will likely depend on the direction indicated by the flavour anomalies.

\section*{Acknowledgements}
The author would like to thank the conference advisory committee for the kind invitation, the local committee
for the perfect conference organisation and \"OBB Nightjet for providing
a carbon-neutral alternative to flying.
Many thanks to
Bo\v{s}tjan Golob,
Neville Harnew,
Niels Tuning
and Gerhard Raven
for helpful comments on the manuscript.
\addcontentsline{toc}{section}{References}
\bibliographystyle{LHCb-10}
\bibliography{LHCb-PAPER,LHCb-CONF,LHCb-DP,exp,theory,local,LHCb-TDR}
\end{document}